# THE TWO PARADIGMS OF SOFTWARE DESIGN

Paul Ralph

Lancaster University

paul@paulralph.name | http://paulralph.name

Abstract

*The dominant view of design in information systems and software engineering, the Rational Design Paradigm, views software development as a methodical, plan-centered, approximately rational process of optimizing a design candidate for known constraints and objectives. This paper synthesizes an Alternative Design Paradigm, which views software development as an amethodical, improvisational, emotional process of simultaneously framing the problem and building artifacts to address it. These conflicting paradigms are manifestations of a deeper philosophical conflict between rationalism and empiricism. The paper clarifies the nature, components and assumptions of each paradigm and explores the implications of the paradigmatic conflict for research, practice and education.*

# 1 INTRODUCTION

Numerous unresolved arguments pervade the discourse on design, information systems development (ISD) and software engineering (SE). Researchers and practitioners disagree on the relative merits of agile and plan-driven methods (Boehm & Turner, 2003), and even on the value of methodicalness itself (Truex et al., 2000) for designing software. Enormous effort has been expended devising more than 1000 software development methods (Jayaratna, 1994), yet many believe methods are rarely used (e.g. Mathiassen & Purao, 2002; Parnas & Clements, 1986). Some argue that system requirements do or should drive the design process (e.g. Kruchten, 2003) while others argue that system requirements are a dangerous illusion (e.g. Ralph, 2012). Some treat design as solving a given problem (Simon, 1996) while others see problem solving and problem framing as fundamentally entangled (Schön, 1983). In practice, managers often attempt to drive projects through budgets, schedules, milestones and lifecycle models, creating conflict with developers and creative professionals who feel unable to provide accurate time estimates (Beck, 2005; Molokken & Jorgensen, 2003).

All of these conflicts may be symptoms of the same underlying phenomenon: the bifurcation of the design discourse into two, possibly incommensurable paradigms (Dorst, 1997; 2006; Dorst & Dijkhuis, 1995). The Rational Design Paradigm is consistent with how many engineers see design – as a methodical, plan-centered, approximately rational process of optimizing a design candidate for known constraints and objectives. The Alternative Design Paradigm is consistent with how many product and industrial designers see design – as an amethodical, improvised, emotional process of simultaneously framing the problem and developing solution artifacts for an unstable, ambiguous context.

Brooks (2010) identified three "formulations" of the Rational Paradigm:

1. the mechanical-engineering view of design as a methodical, orderly process, as described by Pahl and Beitz (1996);
2. the artificial-intelligence view of design as a search for satisfactory alternatives given goals and constraints, by a designer exhibiting "procedural rationality", as formulated by Simon (1996);
3. the managerial view of design as a sequence of weakly-coupled phases, i.e., the Waterfall Model (Royce, 1970).

Similarly, at least three formulations of the Alternative Paradigm are evident:

1. Reflection-in-Action (RiA) – the view of designer as a "reflective practitioner" alternating between problem framing, adjusting a design concept and evaluating the adjustment's consequences (Schön, 1983);
2. the view of the designer as a creative agent whose attention oscillates between an ill-defined problem concept and a tentative solution concept (coevolution), gradually developing an understanding of both (Cross et al., 1992; Dorst & Cross, 2001);
3. the view of design as a political process characterized by interpersonal conflicts, disagreement over goals, politicking and the supremacy of emotional considerations over efficiency (cf. Kling, 1980; Levina, 2005).

However, the Rational Paradigm entails more than just methodicalness, rationality, goal-directed search or the Waterfall Model, and the Alternative Paradigm entails more than reflection, creativity, coevolution and politics. While many have contributed to clarifying the two paradigms (e.g. Brooks, 2010; Dorst, 1997; 2006; Dorst & Dijkhuis, 1995; Schön, 1983), the exact nature and composition of the two paradigms remains ambiguous. This ambiguity furthermore hinders analysis of how paradigmatic differences affect research, practice and education. Consequently, the purpose of this paper is as follows.

> ***Purpose:*** *The purpose of this paper is to clarify the nature and composition of the Rational Design Paradigm and its alternative.*

Here, a *paradigm* is a worldview underlying a theoretical discourse (Kuhn, 1996) and *design* refers to an agent creating a specification of an object to accomplish goals in an environment using a set of primitive components and subject to constraints (Ralph & Wand, 2009). Furthermore, this paper is primarily concerned with innovative, rather than routine, design of software and information systems.

The paper is organized as follows. Section 2 describes the epistemological conflict between rationalism and empiricism. Section 3 clarifies the Rational Design Paradigm (based on rationalism) and synthesizes the "Alternative Design Paradigm" (based on empiricism). The following section examines each paradigm's evaluation and justification (§4). Section 5 summarizes the paper's contributions and limitations, the practical implications of the paradigmatic conflict and avenues for future research.

## 2    DESIGN EPISTEMOLOGY – RATIONALISM VS. EMPIRICISM

Brooks (2010) argues that the Rational Paradigm rests on the philosophy of rationalism. Broadly speaking, *rationalism* is the view that knowledge derives primarily from reason and intuition, in contrast to *empiricism*, the view that knowledge derives primarily from sensory experience. This section provides a brief summary of the dispute between rationalism and empiricism; see Markie (2008) for a more comprehensive discussion.

Technically speaking, one is a Rationalist if one adopts any of three positions:

1. "The Intuition/Deduction Thesis: Some propositions in a particular subject area, S, are knowable by us by intuition alone; still others are knowable by being deduced from intuited propositions."

2. "The Innate Knowledge Thesis: We have knowledge of some truths in a particular subject area, S, as part of our rational nature."

3. "The Innate Concept Thesis: We have some of the concepts we employ in a particular subject area, S, as part of our rational nature." (Markie, 2008)

In other words, some rationalists believe that we are born with certain knowledge (thesis 2) or concepts (thesis 3), which may later be revealed by circumstance. For example, the universal grammar theory of linguistics posits that certain grammatical structures and the capability to learn grammar are hardwired into humans from birth (Cook & Newson, 1996). Of course, grammatical knowledge and concepts will only manifest in babies that are exposed to language.

Meanwhile, rationalists who adopt the Intuition/Deduction Thesis (IDT) believe that some truths are knowable only through intuition or deduction from other intuited truths – they argue that intuiting is a form of rational thought. Many examples from mathematics seem consistent with this view: counting, prime numbers, the transitive property, all appear to spring from intuition and deduction rather than sensory experience. Similarly, in situations where the same proposition is examinable by both intuition/deduction and sensory experience, some rationalists believe that knowledge gained by intuition and deduction is superior to that gained by sensory experience.

In contrast, one is an empiricist if one adopts the following thesis:

4. "*The Empiricism Thesis:* We have no source of knowledge in S or for the concepts we use in S other than sense experience." (Markie, 2008)

In embracing the empiricism thesis, the empiricist rejects the credibility of intuition, innate knowledge, innate concepts and the superiority of rational inference. Sensory experience is, for an empiricist, the final arbiter of truth. Empiricism does not necessarily reject skepticism – the empiricism thesis says that, to the extent that we have knowledge, it is based on sensory experience; it does not assert that we have knowledge. Critical realism, interpretivism, falsificationism and positivism are all forms of empiricism.

As a person may be simultaneously a rationalist in some domains (e.g., mathematics) and an empiricist in others (e.g., biology), empiricism and rationalism do not inherently conflict. However, in domains where some actors adopt a rationalist stance while others adopt an empiricist stance, the rationalism/empiricism conflict manifests. In SE and ISD, the rationalism/empiricism conflict manifests in at least three ways.

First, consider a situation where clients, users and analysts collaborate to devise a system requirements specification. Assuming the system has a clear, agreed goal, the actors infer a set of structural and behavioral properties the system must exhibit to achieve the goal. Treating the requirements specification as knowledge invokes rationalism, specifically IDT. An empiricist, in contrast, would argue that whether a specific property was necessary for success could only be known by building two systems, identical except for that one property, deploying both and seeing which is successful. Only if the system having the property succeeded while the system missing the property failed could one safely conclude that the property was a requirement. In summary, from a rationalist perspective, requirements are credible as participants know them through intuition; from an empiricist perspective, *a priori* requirements specifications are an illusion (Ralph, 2012).

Second, consider a situation where a designer formulates one or more design candidates. If we accept IDT, the designer can formulate good designs by deducing them from the requirements specification and other intuited knowledge. Design thinking is therefore analogous to mathematical thinking: the specification of the problem (requirements) contains the solution; problem solving is therefore reduced to a process of symbol manipulation (cf. Galle, 2009). An empiricist would reject this on the grounds that solution knowledge is not contained in a requirements specification not only because the requirements are illusory (above) but also because solution knowledge is available only through sensory experience, not symbol manipulation.

Third, consider a situation where a designer evaluates a model or prototype of a system. Based on IDT, the designer can discern, by intuition and deduction (from the design and requirements specification) whether the system will achieve its goals. When Simon (1996) imagines the designer exploring the space of possible alternatives, searching for a satisfactory one, he implies that the designer can intuitively tell which designs are satisfactory. An empiricist would reject this and argue that the only way to know whether a design is satisfactory would be to build it, deploy it, and observe its effects (cf. Brooks, 2010).

Having clarified the debate between rationalism and empiricism in this section, the following section elaborates Rational and Alternative Paradigms, which are based on rationalism and empiricism, respectively.

# 3 DECONSTRUCTING THE RATIONAL DESIGN PARADIGM

## 3.1 *Defining the Rational and Empirical Paradigms*

As the Rational Paradigm is both multifaceted and socially constructed, it is difficult to define precisely. Memetics, the study of memes, may help mitigate this difficulty. Dawkins (1989) coined the term *meme* (rhymes with gene) as a "noun that conveys the idea of a unit of cultural transmission, or a unit of imitation" (p. 192). Memes include beliefs, theories, methods, models, philosophies and best practices. Memes may interact and reinforce each other, e.g., the belief that requirements can be stated upfront justifies fixed-price/-schedule contracts and adopting fixed-price/-schedule demands upfront requirements definition. A collection of mutually-reinforcing memes form a complex of memes or *memeplex*. By conceptualizing the Rational Paradigm as a memeplex, it can be defined by its constituent memes and the epistemological position that unites them.

> *The Rational Design Paradigm*: a collection of mutually-reinforcing, design-related memes (including Technical Problem-Solving, plan-driven methods and the Systems Development Lifecycle), which implicitly invoke the Intuition-Deduction Thesis.

This formulation suggests an "Alternative Design Paradigm" comprising alternatives to Rational memes, united by an empiricist epistemological position.

> *The Alternative Design Paradigm*: a collection of mutually-reinforcing, design-related memes (including Reflection-in-Action, amethodical development and Sensemaking-Coevolution-Implementation Theory) which implicitly invoke the Empiricism Thesis.

Tables 1 and 2 summarize the dimensions and assumptions of the Rational and Alternative Paradigms. The remainder of this section discusses core Rational and Alternative memes, and their respective invocations of IDT and Empiricism.

| Dimension | Rational Paradigm | Alternative Paradigm |
|---|---|---|
| epistemology | rationalism[1] | empiricism[1] |
| model of design | Technical Problem-Solving[2] | Reflection-in-Action[3] |
| theory of action | planning model[4] | improvising model[4] |
| development methods | plan-driven[5] | amethodical[6] |
| development process theories | FBS[7], SDLC[8] | SCI[9], Boomerang[10] |

Table 1.     Core Components of the Rational and Alternative Paradigms

Notes: [1](Markie, 2008). [2](Schön, 1983). [3](Simon, 1996). [4](Suchman, 1987). [5](Jacobson et al., 1999; Kruchten, 2003). [6](Truex et al., 2000). [7](Gero, 1990; Gero & Kannengiesser, 2004). [8](Royce, 1970). [9](Ralph, 2013). [10](Stacey & Nandhakumar, 2008).

| Assumption | Rational Paradigm | Alternative Paradigm |
|---|---|---|
| design agent(s) | procedurally rational | emotional and creative |
| design artifact(s) | predictable | complex |
| problems/goals | known, clear and agreed | unknown, ambiguous or conflicting |
| requirements | knowable, drive the design process | illusory; insufficient to drive design |
| analysis, design, programming | loosely coupled | tightly coupled |
| environment | relatively stable | relatively unstable |
| design process | methodical | amethodical |

Table 2.     Assumptions of the Rational and Alternative Paradigms

### 3.2  Technical Problem-Solving vs. Reflection-in-Action.

The foundations of the Rational Paradigm were independently developed in computer science by Herbert Simon, Allen Newell and their collaborators (cf. Newell & Simon, 1972; Simon, 1996) and engineering by Gerhard Pahl, Wolfgang Beitz and their collaborators (cf. Pahl & Beitz, 1996). Design professionals are modeled as rational agents attempting to optimize a design candidate for known constraints and objectives. Where the problem space is large enough to make optimization is intractable, given the designer's limited processing power, the designer will "satisfice" or "find decisions that are good enough" using heuristic search

(Simon, 1996, p. 27). Design agents therefore exhibit "procedural rationality", i.e., although they may not identify the optimal solution, they act rationally given their limited processing power. Schön (1983) called this view "Technical Problem-Solving" (TP-S).

Building on empirical studies of professional practice, Schön (1983) devised Reflection-in-Action (RiA), an alternative to TP-S. RiA models design as a reflective conversation where designer alternates between framing (conceptualizing the problem), making moves (where a move is a real or simulated action intended to improve the situation) and evaluating those moves. Groups collectively reflect in action by externalizing their cognition using boundary objects (Levina, 2005). Boundary objects, including diagrams and prototypes, are simultaneously robust enough to maintain their identities and flexible enough to serve multiple parties (cf. Bergman et al., 2007).

TP-S adopts IDT by making the three assumptions discussed in Section 2: that requirements are credible, that designs may be deduced from requirements, and that design quality can be inferred by inspection. RiA, in contrast, adopts the empiricism thesis in assuming that designers co-construct the problem and the solution rather than deducing a solution from a given problem specification. However, RiA's "making moves" and "evaluating moves" steps have two interpretations. If making moves involves changing mental, virtual or physical models of the design artifact, and evaluating moves involves mental simulations, then RiA invokes IDT. However, if making moves includes changes to the design artifact itself and evaluating moves entails examining real consequences, RiA invokes empiricism. In domains including architecture and mechanical engineering, the rationalist interpretation of RiA would likely be more reasonable. However, given the virtual nature of software and information systems, this paper adopts the empiricist interpretation.

*3.3  Planning vs. Improvising*

Cognitive science is dominated by two competing views of human action: the planning model and the improvising model (Suchman, 1987). According to the planning model, planning is a prerequisite to acting, actors understand their actions and progress in terms of the plan, and good planning is a primary determinant of success. In contrast, the improvising model posits that "the organization of situated action is an emergent property of moment-by-moment interactions between actors, and between actors and the environments of their action" (Suchman, 1987, p. 179) while "plans are representations, or abstractions over action" (p. 186). To be clear, both models involve planning, but while planning is the basis of action in the planning model, plans in the improvising model are merely abstractions or post-hoc rationalizations.

In SE and ISD, the planning model underlies stage-gate or lifecycle views of the design process (Boehm, 1988; Royce, 1970), plan-driven methods (Jacobson et al., 1999; Kruchten, 2003), and the perceived need for significant upfront analysis. Meanwhile, the improvisation model underlies teleological views of the design processes (Ralph, 2013; Stacey & Nandhakumar, 2009), agile methods (Beck, 2005; Schwaber, 2004) and the desire to postpone unnecessary decisions.

Neither the planning model nor the improvisation model are inherently associated with rationalism or empiricism; however, when applied to SE and ISD, they become entangled with alternative epistemologies. The planning vs. improvising debate manifests in SE/ISD in two senses.

First, the utility of planning for an SE/ISD project depends on the planner's ability to answer questions including *what tasks does this project entail, how long will each task take, what resources will be needed, what risks are involved (and what is the probability of each)* and *what phases will the project pass through?* Invoking IDT, a rationalist would argue that the planner may answer these questions using intuition and deduction; however, an empiricist would argue that to the extent these questions could be answered at all, they could only be answered by first building the system. In the SE/ISD domain, therefore, the planning model is entangled with rationalism while the improvising model is entangled with empiricism.

Second, in TP-S and many software development methods, 'design' is seen as a kind of planning, specifically, converting a requirements specification into a plan for constructing an artifact. RiA, in contrast, views design as a kind of improvising where a designer simultaneously refines a problem and a solution candidate. The view of design as planning is entangled with rationalism – it only makes sense if requirements are given and credible, and if the quality or effectiveness of the artifact is evident from its blueprints. If, from an empiricist perspective, we reject requirements as absent and assume that design quality is only discernible by building the system, the view of designing as planning appears less useful than the view of the designer as improvising. In this way, the improvising model is consistent with empiricism in that design is seen as an interaction between an actor and its environment; a design project is seen as a system of inquiry where the designer generates solutions by building artifacts in the world rather than planning them in the mind.

*3.4 System Development Methods*

A *System Development Method* (SDM) is a collection of prescriptions concerning how to build a software or information system effectively. Many modern SDMs retain elements of the Waterfall model. For example, Extreme Programming (Beck, 2005) involves running through Waterfall phases in linear, time-boxed iterations (Figure 1), while the Unified Process (UP) (Jacobson et al., 1999) adopts them as parallel "disciplines" within an alternative phase model (Figure 2). Even Boehm's (1988) spiral model is composed primarily of Waterfall phases reorganized into a spiral with added risk analysis.

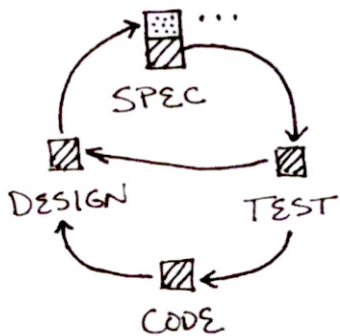 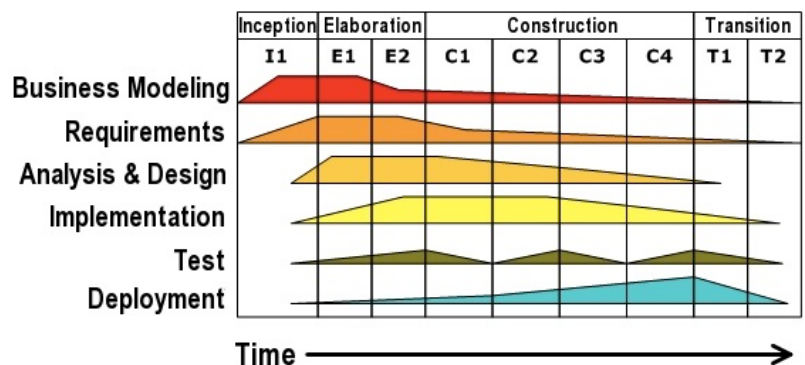

*Fig. 1. XP's "Pull Model of Development" (Beck 2005)*

*Fig. 2. Unified Software Process Phase Model (adapted from Wikimedia Commons)*

Modern SDMs are often divided into two categories – *plan-driven* SDMs, including Waterfall and UP, recommend extensive planning, up-front analysis, and a proactive mindset; *Agile* SDMs (Beck et al., 2001), including Extreme Programming (Beck, 2005) and Lean (M. Poppendieck & Poppendieck, 2003), emphasize people (over process), rapid prototyping and a reactive mindset. However, not all projects use a defined method (e.g. Zheng et al., 2011) – in fact, *amethodical development,* which "implies management and orchestration of systems development without a predefined sequence, control, rationality, or claims to universality", may be the norm (Truex et al., 2000, p. 54).

Plan-driven methods are deeply intwined with the planning model of cognitive science (above). Just as the planning model posits that plans are the basis for human action, plan-driven methods employ plans as the basis for managing, controlling and understanding progress. Plan-driven methods invoke IDT by assuming that project actors can, through intuition and deduction, anticipate the problems the project will encounter, the phases it will pass through and any changes to its context. Plan-driven methods implicitly or explicitly assert that good planning is an important driver of project success.

In contrast, amethodical development is primarily concerned with improvisation (Truex et al., 2000; Zheng et al., 2011), consistent with the improvising model of cognitive science. Moreover, amethodical development invokes the empiricism thesis by rejecting the proposition that project actors can anticipate future problems, phases and changes. As actors will not know what will happen until it does and is available to sensory perception, the logical approach would be to cultivate the capacity to adapt and respond quickly to unexpected changes, i.e., to maximize agility.

In principle, Agile methods value "responding to change over following a plan" (Beck et al., 2001), which suggests consistency with the empiricism thesis. However, in practice, the situation is complicated by numerous factors:

1. Many purportedly Agile methods have been proposed (Abrahamsson et al., 2009).
2. Little research has directly addressed the relationship between ostensibly Agile practices and the actual responsiveness of the teams that use them (Lee & Xia, 2010).
3. Specific agile methods may combine empiricist and rationalist logics. For example, Extreme Programming (Beck, 2005) prescribes both: 1) release and iteration planning each comprising three specific phases, which seems rationalist, and 2) continuously releasing small functionality increments to get stakeholder feedback, which seems empiricist.

A comprehensive review of Agile methods and their individual invocations of rationalism and empiricism is beyond the scope of this paper. However, for our purposes, it is sufficient to point out the plan-driven methods invoke primarily rationalist logics, amethodical development invokes primarily empiricist logics, and Agile methods may combine the two.

### 3.5   *Software Development Theories*

A *process theory* is an explanation of how and why an entity changes and develops (Van de Ven, 2007). Unlike a SDM, which attempts to describe *one good way* of developing software, a software process theory attempts to explain *all the ways* of developing software (Vermaas & Dorst, 2007). That is, a software process theory aims to illuminate the phenomena fundamental to creating software, not make prescriptions. Despite many calls for developing theories of software development (e.g. Jacobson & Meyer, 2009; Johnson et al., 2012; Ralph et al., 2013), no widely accepted theory of the software development process is evident (Johnson et al., 2012). Instead, it seems that some variant of the Waterfall Model has been implicitly co-opted as the dominant software development process theory (Brooks, 2010).

Royce (1970) initially presented Waterfall as a method, which he believed had "never worked on large software development efforts" (p. 335). Minor variations on the Waterfall model, some including cycling and iteration (Figure 3), came to be known as the Systems Development Lifecycle (SDLC) (Gladden, 1982). Despite, many academics and practitioners criticizing SDLC as ineffective or unrealistic (e.g. Beck, 2005; Boehm, 1988; Curtis et al.,

1992; Gladden, 1982; McCracken & Jackson, 1982), many others began treating SDLC as a process theory. For example, Fitzgerald (2006) states that "in conventional software development, the development lifecycle in its most generic form comprises four broad phases: planning, analysis, design, and implementation" (p. 3). Similarly, Ewusi-Mensah (Ewusi-Mensah, 2003) says that "regardless of the particular process model … every software project will feature: (1) the requirements-definition and functional-specification phase; (2) the design phase; … (3) the implementation; … and (4) the installation, operation, and maintenance phase" (p. 51). Additionally, Laudon et al. (2009) state that "systems development … consist[s] of systems analysis, systems design, programming, testing, conversion and production and maintenance … which usually take place in sequential order". Moreover, traditional SDLC phases are explicitly adopted by the IEEE Guide to the Software Engineering Body of Knowledge (Bourque & Dupuis, 2004) and implicitly adopted by numerous standards (e.g. Fujitsu et al., 2012; IEEE, 1998; *Systems and software engineering -- Life cycle processes --Requirements engineering*, 2011).

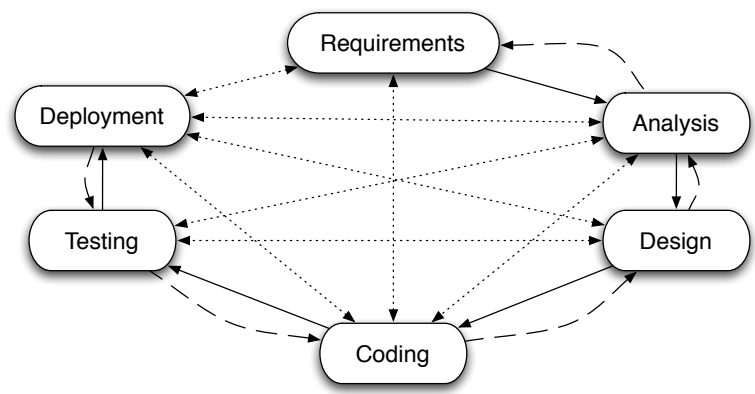

*Fig. 3.* *SDLC* (adapted from Royce, 1970).

Note: Solid lines indicate the original, no-backtracking version. Solid lines plus dashed lines show the original version with backtracking. Solid, dashed and dotted lines indicate the clique version (see below).

The SDLC that is often explicitly or implicitly invoked as a de facto process theory does not necessarily feature the linear progression for which Waterfall is named. Rather, this new *clique*-SDLC is defined by categorizing development activities into distinct, separable classes including analysis, design, coding and testing. These activity classes, not linear progression, are the core proposition of the contemporary SDLC.

An alternative to SDLC is Sensemaking-Coevolution-Implementation Theory (SCI). SCI (Ralph, 2013) posits that complex software systems are produced primarily through its three titular activities (Figure 4, Table 3). Simply, where SDLC organizes development activities into classes including analysis, design, programming and testing, SCI organizes development into sensemaking, coevolution and implementation. Sensemaking may include interviewing stakeholders, writing and organizing notes, reading about the domain, investigating technologies for use in the project, sharing insights among team members, and acceptance testing. Implementation may include coding, managing the codebase, writing documentation, unit testing, and debugging. Coevolution, in contrast, does not directly map to a variety of well-known software engineering activities. Rather, it refers to the mutual exploration of the context and design alternatives, as seen in meetings where participants oscillate between conceptual models and design models, changes in one triggering changes in the other and vice versa.

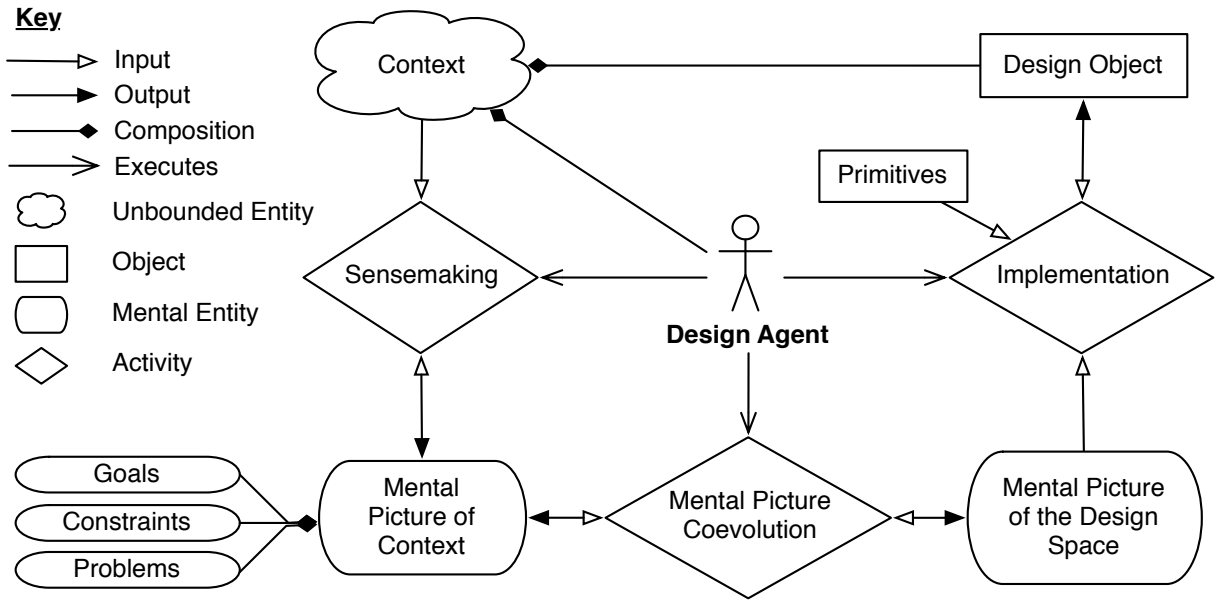

*Fig. 4. Sensemaking-Coevolution-Implementation Theory* (adapted from Ralph, 2013)

| Concept / Activity | Meaning |
| --- | --- |
| Constraints | the set of restrictions on the design object's properties |
| Design Agent | an entity or group of entities capable of forming intentions and goals and taking actions to achieve those goals and that specifies the structural properties of the design object |
| Design Object's Environment | the totality of the surroundings where the design object exists or is intended to exist |
| Design Agent's Environment | the totality of the surroundings of the design agent |
| Design Object | the thing being designed |
| Goals | optative statements about the effects the design object should have on its environment |
| Mental Picture of Context | the collection of all of the design agent's beliefs about its and the design object's environments |
| Mental Picture of Design Object | the collection of all of the design agent's beliefs about the design object |
| Primitives | the set of entities from which the design object may be composed |
| Sensemaking | the process where the design agent perceives its and the design object's environments and organizes these perceptions to create or to refine the mental picture of context |
| Coevolution | the process where the design agent simultaneously refines its mental picture of the design object, based on its mental picture of context, and the inverse |
| Implementation | the process where the design agent generates or updates the design object using its mental picture of the design object |

*Table 3.        SCI Theory Concepts and Relationships* (adapted from Ralph, 2013)

In distinguishing coevolution from evolution, SCI further posits that software development entails two iterative loops. In the fast, coevolutionary loop, changes in the mental picture of the design object trigger changes to the mental picture of the context, and vice versa, typically over minutes or hours. In the slow, evolutionary (prototyping) loop, changes to the actual design object trigger changes in the actual context, and vice versa, typically over weeks and months.

SDLC and SCI invoke opposite epistemological theses. SDLC invokes IDT by making the same assumptions discussed in Section 2: that requirements are credible, that designs may be deduced from requirements, and that design quality can be inferred by inspection. In contrast, SCI invokes the empiricism thesis by positing that the design agent forms beliefs about the project context by perceiving the context, and evaluates design quality be deploying the software and observing its effects.

Of course, SDLC and SCI are not the only theories pertaining to the software development process. The Function-Behavior-Structure Framework of engineering design (FBS) posits that a design agent manipulates models of the functions (goals), behaviors and structure of an artifact to produce a detailed specification suitable for informing construction (Gero, 1990; Gero & Kannengiesser, 2004). FBS can be applied to software engineering (Gero & Kannengiesser, 2007), where it clearly invokes a rationalist epistemology by making similar assumptions to SDLC and TP-S; in fact, Kruchten (2005) illustrates how SDLC can be seen as a specific instance of FBS. In contrast, the Boomerang theory of the video game development process (Stacey & Nandhakumar, 2008) posits that game development is driven through play-testing, which not only identifies coding errors but also triggers redesign and even reconceptualization of projects and their goals. Boomerang clearly invokes an empiricist logic wherein actors primarily derive knowledge from the sensory experience of play-testing.

# 4    EVALUATING ACCURACY

Exploring the Rational and Alternative Paradigms raises an obvious question: *which paradigm is a more accurate depiction of software development,* or perhaps, *under what circumstances is each paradigm accurate?* As these are difficult and controversial questions, this sections builds up to them by first exploring less controversial aspects of the two paradigms.

## *4.1   Relative Dominance and Justification*

First, the Rational Paradigm is the dominant paradigm in IS and SE (Brooks, 2010; Schön, 1983), as seen in textbooks (e.g. Laudon et al., 2009), model curricula (Joint Task Force on Computing Curricula, 2004; Topi et al., 2010), standards (e.g. Fujitsu et al., 2012; IEEE, 1998; *Systems and software engineering -- Life cycle processes --Requirements engineering*, 2011), official body of knowledge compilations (e.g. Bourque & Dupuis, 2004), Wikipedia articles (e.g., the ubiquitous [Software Development Process Template](#)), the elevation of SDLC to a de facto process theory (discussed above), and even popular conceptions of the scientific process (e.g. Campbell & Stanley, 1963; Trochim, 2001; Yin 2008). The Rational Paradigm also pervades our research; for example, Hevner et al. (2004) – channeling TP-S and Simon (1996) – state that "Design is essentially a search process to discover an effective solution to a problem" (p. 88). Presumably, no reviewer pointed to observations of professional designers that question the search metaphor (Cross et al., 1992), no reviewer remarked that designers often encounter problematic situations characterized by many stakeholders with myriad, vague or conflicting views, rather than a specific problem (Checkland, 1999), no reviewer generally demanded that Hevner et al. justify this statement. The controversial nature of such statements is obscured by the dominance of the Rational Paradigm.

Second, several reasons are evident for the Rational Paradigm's continued dominance. It came first. It was developed by widely respected, influential academics. And, to the best of my knowledge, no comprehensive alternative was previously proposed. (Brooks suggests replacing Waterfall with the Spiral Model (Boehm, 1988) but the Rational Paradigm is substantially broader than either of these models.) Furthermore, a host of social and cognitive factors including the Semmelweis reflex (Shore, 2008) and status quo bias (Samuelson & Zeckhauser, 1988) inhibit paradigm shifts.

Furthermore, one important aspect of memes is that they may have properties, other than truth, that increase their propagation (Dawkins, 1989). The Rational Paradigm exhibits many such properties. Carefully planning projects seems intuitively more *mature* than improvising, cowboy coding or "flying by the seat of your pants". A rational designer, systematically searching a solution space, optimizing against hard constraints, seems more *professional* than an emotional creative imagining solutions from thin air. Dividing a team into clearly identifiable job roles – analyst, programmer, tester – seems more *organized* than calling everyone "Team Member" and allowing them to self-select their work. Finally, when confronted with stakeholders who cannot agree on goals, analysts who cannot specify requirements, designers who cannot deduce solutions and testers who cannot conclusively prove the system works, many feel that they *should* be able to, suggesting that even if the Rational Paradigm does not accord with every single observation, it is a kind of ideal to which we should strive (cf. Parnas & Clements, 1986).

Third, proponents of the Rational and Alternative Paradigms employ different forms of justification; unsurprisingly, the Rational Paradigm relies on a rationalist justification while the Alternative Paradigm relies on empirical justification. For example, Simon (1996) simply lays out TP-S as a way of thinking about design, relying on its intuitive resonance for justification, while Schön (1983) justifies RiA based on analysis of extensive observations of professional designers. Similarly, Gero and his colleagues justify FBS based on conceptual evaluations (Gero, 1990; Gero & Kannengiesser, 2004; cf. 2007) while Stacey and Nandhakumar (2008) justify Boomerang based on a multiple-case study of game developers and Ralph (2010a) justifies SCI based on a questionnaire study of software developers.

### *4.2 Differing views of the two paradigms*

Returning to the question of which paradigm is more appropriate for the software development domain, we are left with a potentially unsatisfactory answer. As the Rational Paradigm relies on conceptual justification, its proponents may dismiss empirical evidence contradicting its claims; similarly, as the Empirical Paradigm relies on empirical justification, its proponents may dismiss conceptual critiques. This epistemological conflict undermines consensus.

More specifically, the full-fledged rationalist may or may not accept either paradigm based on their relative intuitive appeal. Meanwhile, the full-fledged empiricist will clearly reject the Rational Paradigm as it lacks empirical support (Bansler & Bødker, 1993; Brooks, 2010; Cross et al., 1992; Zheng et al., 2011). However, empiricists' views of the Alternative Paradigm will vary depending on the component in question, the evidence available and the empirical standard applied. Yet others may view the Rational Paradigm more as an ideal process which may differ from actual processes, especially for inexperienced or less capable teams.

The fundamental rationalist critique of the Alternative Paradigm is that employing a more systematic, methodical, logical process will intuitively improve outcomes (Parnas, 2003; 2009; Parnas & Clements, 1986). The fundamental empiricist critique of the Rational Paradigm is that real designers do not follow systematic, methodical logical processes and there is no empirical evidence that forcing them to will improve outcomes (Brooks, 2010; Cross et al., 1992). However, the available empirical evidence paints a more Escheresque picture.

### 4.3 Summary of Empirical Evidence

Concerning TP-S and RiA, TP-S assumes that design involves solving given problems. However, observational studies demonstrate that, in practice, problem framing and solution design are interconnected and simultaneous (e.g. Cross et al., 1992; Dorst & Cross, 2001). Schön (1983) found that a designer "does not keep means and ends separate, but defines them interactively as he frames a problematic situation. He does not separate thinking from doing" (p. 69). In contrast, RiA is explicitly based on the empirical foundation provided by Schön's (1983) case studies. Furthermore, RiA was extended for team design based on Levina's (2005) field studies. Meanwhile, the core TP-S assumption that problems/goals are known unambiguous, and agreed on by stakeholders is incommensurate with empirical evidence of goal conflict in design projects (cf. Checkland, 1999). However, one protocol study found that both TP-S and RiA can be used to describe design activity in different circumstances (Dorst & Dijkhuis, 1995).

Concerning theories of human action, Suchman's (1987) comprehensive comparison of the planning and improvising views concludes that action is inherently improvised. Numerous studies of software development, specifically, support the key role of improvisation (e.g. Arvola & Artman, 2007; Magni et al., 2008; Stacey & Nandhakumar, 2009; Zheng et al., 2011).

Concerning methodicalness, the situation is more complex. Some evidence suggests that process maturity, which is closely related to methodicalness, is associated with numerous benefits including higher morale and productivity (Herbsleb et al., 1997). However, others have found that shifting to Agile methods (less methodical) reduces project failures (Group, 2006) and increases developer productivity (Cardozo et al., 2010). Additionally, some teams clearly act amethodically (Baskerville & Pries-Heje, 2004; Baskerville et al., 1992) and they can also be successful (Zheng et al., 2011). Meanwhile, SDMs are neither effectively nor extensively used (Avgerou & Cornford, 1993; Bansler & Bødker, 1993; Dobing & Parsons, 2006; Whitley, 1998) and when they are used, they are not used as intended (Mathiassen & Purao, 2002). Rather, "traditional IS development methodologies are treated primarily as a necessary fiction to present an image of control or to provide a symbolic status, and are too mechanistic to be of much use in the detailed, day-to-day organization of systems developers' activities" (Nandhakumar & Avison, 1999, p. 176). Moreover, little or no credible comparative testing of SDMs has been done (Wynekoop & Russo, 1997), possibly due to the intense methodological challenges of such work (Ralph, 2010b). In summary, team methodicalness and method use vary, and their impacts are not well understood.

Concerning process theories, an extensive search using Academic Search Premier and Google Scholar (Query 1) produced only one empirical study supporting SDLC. Palvia and Nosek (Palvia & Nosek, 1990) evaluated SDLC against a prototyping methodology using a survey;

however, the instrument explicitly assumed that SDLC describes all software development, creating a circular argument. Similar queries found no empirical studies of FBS. In contrast, a survey of more than 1400 software developers (Ralph, 2010a) found that SCI more accurately represented software development than SDLC or FBS and a multiple case study of three game development teams supported Boomerang (Stacey & Nandhakumar, 2008; cf. 2009)

> ***Query 1:*** *("Waterfall Model" OR SDLC) AND (experiment OR "laboratory study" OR "case study" OR "field study" OR survey OR "variance model" or "econometric analysis")*

In summary, while the Alternative Paradigm is generally more consistent with the empirical evidence than the Rational Paradigm, more research is clearly needed.

# 5   CONCLUSION

This paper makes three contributions. First, it extends Brooks' (2010) identification of the Rational Design Paradigm by clarifying its nature (as a memeplex) and enumerating many of its component memes. Second, it exposes the differing epistemological positions underlying many conflicts in the SE and ISD discourse. Third, it organizes alternatives to each Rational meme into the Alternative Design Paradigm.

These contributions should be considered within several limitations. First, the paper does not attempt a comprehensive account of every possibly aspect of the two paradigms as, given their complexity, it would obfuscate the core issues. Second, a comprehensive discourse analysis organizing the SE and ISD literature according to Rational and Alternative memes is beyond the scope of this paper (but a potentially-useful avenue for future research). Third, the above analysis may be unconsciously biased by the empiricist leanings of the author. Fourth, paradigms are not definitively right or wrong, true or false, or more or less correct. However, a paradigm may be applied in situations where it is inappropriate, and adopting one paradigm or the other will have diverse consequences for research, practice and education.

The dominance of the Rational Paradigm affects research in many ways. For example, the design science research approach is heavily influenced by TP-S – Hevner et al. (2004) cite Simon 13 times, conceptualize "the design science paradigm" as "fundamentally a problem-solving paradigm" (p. 76) and advise research to "design as a search process" (p. 88) and evaluate artifacts based on requirements. Therefore, much of the advice for conducting and evaluating design science research may not apply in situations where the researcher chooses his own problem, where the design space is too poorly understood to "search" or where requirements are unknown or unclear. However, perhaps due to the Rational Paradigm's dominance, these limitations are rarely discussed in the design science literature (Baskerville et al., 2009 for notable exceptions; cf. Sein et al., 2011).

Furthermore, much research involves developing new and improved software development methods, tools and practices. If researchers are unknowingly immersed in the Rational Paradigm, they may create methods, tools and practices based on unstated rationalist assumptions, limiting their applicability or even making them counterproductive. For instance, the Unified Process organizes work into six "engineering disciplines" – "business modeling", "requirements", "analysis and design", "implementation", "testing" and "deployment" (Jacobson et al., 1999). By assigning, for example, "implementation" work to the "programmer" role and "analysis and design" work to the "architect" role, UP not only assumes that design and programming are separable but also posits that separating them is a good idea. In domains where design and programming are inextricably entangled, forcing them into separate roles will be counterproductive. Again, perhaps due to the Rational Paradigm's dominance, this sentiment is entirely absent from the literature on UP and the closely-related Rational Unified Process (Kruchten, 2003) and Agile Unified Process (Ambler & Lines, 2012).

Moreover, the implications of the Rational and Alternative paradigms for software development practice extend well beyond practitioners' use of concepts from academia. Brooks (2010) argues that the Rational Paradigm leads project actors "to demand up-front statements of design requirements" and "to make contracts with one another on [this] basis" (p. 33-34). Brooks' argument involves at least three interconnected phenomena – 1) upfront requirements; 2) outsourcing; 3) fixed-price/-schedule contracts. The Rational Paradigm includes the assumptions and beliefs that justify these ideas – 1) systems have known, unambiguous, agreed goals; 2) requirements can be deduced from given goals; 3) analysis activities are loosely coupled to design and implementation activities. The Alternative Paradigm, in contrast, assumes that goals are unclear, requirements are illusory and problem framing is tightly-coupled to problem solving. The intellectual case for outsourcing based on given requirements and fixed-price/-schedule contracts is entwined with the Rational Paradigm whereas the Alternative Paradigm may help explain many of difficulties encountered in global software development (Herbsleb & Moitra, 2001; Holmstrom et al., 2006). Empirically speaking, developers underestimate effort by at least 30-40% on average (Molokken & Jorgensen, 2003) as they rarely have sufficient information to gauge project difficulty (Ralph & Wand, 2009). This contributes to the well-known tension between managers attempting to drive projects through cost estimates and developers who are unable to accurately estimate costs (Beck, 2005).

More generally, IEEE standards (e.g. IEEE, 1998; *Systems and software engineering -- Life cycle processes --Requirements engineering*, 2011), OMG standards (e.g. Fujitsu et al., 2012), the Capability Maturity Model (Herbsleb et al., 1997) – now Capability Maturity Model Integration (Team, 2011) and much of the project management discourse (e.g. Bourque & Dupuis, 2004; IIBA, 2009; Project Management Institute, 2004) – all manifest the assumptions and logics of the Rational Paradigm. The presentation of Rationalist assumptions in the discourse may fuel the impression that "serious", "professional" or "expert" actors do or should embrace the Rational Paradigm; while the Alternative Paradigm may only describe the work of "frivolous", "amateur" or "incompetent" actors. This impression may lead managers, for example, to criticize developer's useful and necessary improvisation as capricious and unprofessional.

The dominance of the Rational Paradigm can also be seen in IS2010 (Topi et al., 2010) and SE2004 (Joint Task Force on Computing Curricula, 2004), the undergraduate model curricula for information systems and software engineering, respectively. Both curricula discuss requirements and SDLC in depth; neither mention reflection-in-action, coevolution, amethodical development or any theories of SE or ISD. Graduates accustomed to given problems and idealized processes may be ill-prepared for the messy, political reality of their first job.

These implications suggest several avenues of future research. First, direct empirical comparisons of SDMs, a methodologically challenging task, are needed. Second, some descriptive research on the extent to which Rational-Model assumptions are met in diverse projects would be beneficial. Finally, the components of the Alternative Paradigm may be clarified and improved. Numerous potential extensions for SCI, in particular, are evident including integrating boundary objects (Bergman et al., 2007), modeling agents as transactive memory systems (Wegner, 1987) and generalizing the theory for multi-agent contexts.

In conclusion, this paper clarifies the nature, components and implications of the Rational Design Paradigm, the dominant view of design in IS and SE. It further synthesizes a comprehensive alternative, clarifying the root of many conflicts in the development and management discourse, and bringing into stark relief the possibility of a paradigm shift in design research, practice and education.